\documentclass[12pt]{article}
\usepackage{a4wide,epsfig}

\newcommand{\sineff}{\sin^2\theta_{\rm eff}^{\rm lept}}

\begin{document}


\title{\vskip-3cm{\baselineskip14pt
    \begin{flushleft}
      \normalsize BI-TP 2005/12\\
      \normalsize SFB/CPP-05-11 \\
      \normalsize TTP05-03 \\
  \end{flushleft}}
  \vskip1.5cm
  Four-Loop Singlet Contribution to the Electroweak\\ $\rho$ Parameter
}
\author{\small Y. Schr\"oder$^a$ and M. Steinhauser$^b$\\
{\small\it $^a$ Fakult\"at f\"ur Physik, Universit\"at Bielefeld,}\\
{\small\it 33501 Bielefeld, Germany}\\
{\small\it $^b$ Institut f{\"u}r Theoretische Teilchenphysik,
  Universit{\"a}t Karlsruhe,}\\
{\small\it 76128 Karlsruhe, Germany}\\
}

\date{}

\maketitle

\thispagestyle{empty}

\begin{abstract}
  We compute the four-loop QCD contribution to the electroweak $\rho$
  parameter induced by the singlet diagrams of the $Z$-boson
  self-energy. The numerical impact on the weak mixing angle and the
  $W$-boson mass is small.

\medskip


\end{abstract}

\newpage


\section{Introduction}

The electroweak $\rho$ parameter as introduced by Veltman~\cite{Veltman:1977kh}
measures the relative strength of the charged and neutral current.
Considering QCD corrections it can be written as
\begin{eqnarray}
  \rho &=& 1 + \delta\rho
  \,,
\end{eqnarray}
with
\begin{eqnarray}
  \delta\rho &=& \frac{\Pi_{ZZ}(0)}{M_Z^2} - \frac{\Pi_{WW}(0)}{M_W^2}
  \,.
\end{eqnarray}
$\Pi_{ZZ}(0)$ and $\Pi_{WW}(0)$ are the transverse parts of the 
$W$- and $Z$-boson self-energies evaluated for vanishing external momentum.
The parameter $\delta\rho$
enters a variety of quantities which are determined from
experiment with an enormous precision.
In particular, it enters the relation between the
$W$-boson mass, $M_W$, the fine structure constant, $\alpha$, the Fermi
constant, $G_F$, and the $Z$-boson mass, $M_Z$, which is given by~\cite{Sirlin:1980nh}
\begin{eqnarray}
  M_W^2 &=& \frac{M_Z^2}{2}\left(1
  +\sqrt{1-\frac{4\pi\alpha}{\sqrt{2}\, M_Z^2 G_F\left(1-\Delta
  r\right)}}
  \right)
  \,.
  \label{eq::MW}
\end{eqnarray}
The quantity $\Delta r$ is conveniently parameterized in the form
\begin{eqnarray}
  \Delta r &=& \Delta\alpha - \frac{c_W^2}{s_W^2} \delta\rho + \Delta r^{\rm rem}
  \label{eq::delr}
  \,,
\end{eqnarray}
with $c_W=M_W/M_Z$ and $s_W^2=1-c_W^2$.
$\Delta\alpha$ contains contributions from light fermions
giving rise to a correction of about 6\%. The leading corrections
proportional to $G_F M_t^2$ are incorporated in $\delta\rho$
and amount at one-loop order 
to roughly $-3$\% whereas the remaining part is small.

Eqs.~(\ref{eq::MW}) and~(\ref{eq::delr}) 
can be used to predict $M_W$ where the formula
\begin{eqnarray}
  \delta M_W &=& \frac{M_W}{2} \frac{c_W^2}{c_W^2-s_W^2}\, \delta\rho
  \,,
  \label{eq::delMW}
\end{eqnarray}
immediately accounts for the dominant shift 
in $M_W$ due to the corrections to the
$\rho$ parameter.
We can also look at the change of the effective leptonic weak
mixing angle, $\sineff$, defined through the coupling of the $Z$-boson
to leptons. The leading universal corrections originating from
$\delta\rho$ can in analogy to Eq.~(\ref{eq::delMW}) be written as
\begin{eqnarray}
  \delta\sineff &=& - \frac{c_W^2 s_W^2}{c_W^2-s_W^2} \, \delta\rho
  \,.
  \label{eq::delsin}
\end{eqnarray}

Currently the uncertainties for $M_W$ and $\sineff$ are given by 
$\delta M_W=34$~MeV and $\delta\sineff=1.7\times
10^{-4}$~\cite{LEPEWWG}, respectively.
However, a future linear collider running at the $Z$-boson pole, the 
so-called Giga-$Z$ option, and around the $W$-pair threshold
might reduce the uncertainties to $\delta M_W=6$~MeV and 
$\delta\sineff=1.3\times 10^{-5}$~\cite{Aguilar-Saavedra:2001rg}.

\begin{figure}[t]
      \leavevmode
      \epsfxsize=14.0cm
      \epsffile[80 400 500 450]{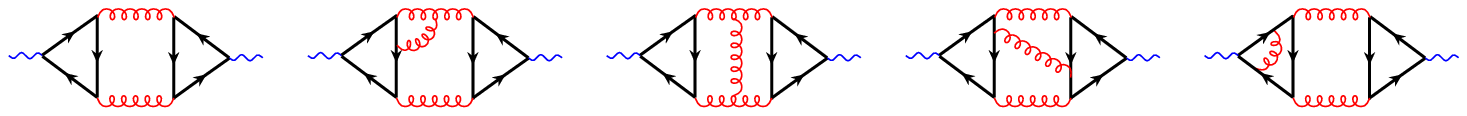}
  \vspace*{-1em}
  \caption{\label{fig::sing}
    Sample three- and four-loop 
    singlet diagrams contributing to the $\rho$ parameter.
    In the fermion loops either top- or bottom-quarks are present.
          }
\end{figure}

The one-loop corrections to $\rho$ have been computed in
1977~\cite{Veltman:1977kh} 
and also the two-loop QCD corrections are known since almost 20
years~\cite{Djouadi:1987gn,Djouadi:1987di,Kniehl:1988ie}. 
Roughly 10 years ago the order $G_F M_t^2 \alpha_s^2$ QCD
corrections~\cite{Avdeev:1994db,Chetyrkin:1995ix}
constituted one of the first applications of the three-loop massive
vacuum integrals.
At three-loop order for the first time a new kind of Feynman graphs
has to be considered, the so-called singlet diagrams 
as shown in Fig.~\ref{fig::sing} which only contribute to the 
$Z$-boson self-energy. They are characterized by the fact that 
in contrast to the non-singlet contribution 
the external $Z$-bosons couple to different fermion lines.
We want to note that the singlet contribution forms a finite 
and gauge independent subset. At three-loop order it completely dominates 
the numerical corrections if the $\overline{\rm MS}$ definition is
adopted for the top-quark mass. In the case of the pole mass
definition the singlet part still amounts to about 30\% of the 
total three-loop contribution.
We want to mention that also
two-loop~\cite{vanderBij:1986hy,Barbieri:1992nz} 
and three-loop
mixed electroweak/QCD~\cite{Faisst:2003px} and even three-loop
pure electroweak corrections~\cite{Faisst:2003px} have been evaluated.
Recently also corrections in the large Higgs boson mass limit have
been considered~\cite{Boughezal:2004ef,Boughezal:2005eb}. For
non-universal corrections to $M_W$ and $\sineff$ we refer
to~\cite{Awramik:2003rn,Awramik:2004ge}.

In this letter we consider the four-loop contribution to the $\rho$
parameter originating from the singlet diagrams. 
In Fig.~\ref{fig::sing} some sample diagrams are shown.
This constitutes one
of the first applications of the four-loop vacuum master integrals
evaluated recently in Ref.~\cite{Schroder:2005va}.


\section{Technicalities}

Since the boson self-energies have to be evaluated for zero external
momentum and only QCD corrections are considered, 
only the axial-vector part of the $Z$-boson correlator gives a
non-zero contribution. Whereas for the non-singlet contribution
the naive anti-commuting definition of $\gamma_5$ can be adopted, 
special care has to be taken in the singlet case.
Actually, the definition of 't~Hooft and Veltman~\cite{'tHooft:1972fi} has to be
adopted and additional counterterms have to be introduced in order to
ensure the validity of the Ward identities.
In the practical calculation we follow Ref.~\cite{Larin:1993tq} and perform the
following replacement in the axial-vector current
\begin{eqnarray}
  \gamma^\mu\gamma^5 &=& 
  \frac{1}{3!} \varepsilon^{\mu\nu\rho\sigma} 
  \gamma_{\nu}\gamma_{\rho}\gamma_{\sigma}
  \,.
\end{eqnarray}
We pull out the $\varepsilon$-tensor from the actual integral and 
consider instead the completely antisymmetrized 
product of the three $\gamma$-matrices which can be written as
\begin{eqnarray}
  \gamma^{[\nu}\gamma^{\rho}\gamma^{\sigma]} &=&
  \frac{1}{2}\left(\gamma^{\nu}\gamma^{\rho}\gamma^{\sigma}
  - \gamma^{\sigma}\gamma^{\rho}\gamma^{\nu} \right)
  \,.
\end{eqnarray}
As a consequence we have to deal with an object with six indices.
Thus, for zero external momentum we obtain 
\begin{eqnarray}
  \Pi_{\rm ZZ} &=& \frac{g_{\mu\mu^\prime}}{4}\Pi_{\rm ZZ}^{\mu\mu^\prime}
  \nonumber\\
  &=& \frac{g_{\mu\mu^\prime}
  \varepsilon^{\mu\nu\rho\sigma}
  \varepsilon^{\mu^\prime\nu^\prime\rho^\prime\sigma^\prime}
  }{144}
  \Pi_{[\nu\rho\sigma][\nu^\prime\rho^\prime\sigma^\prime]}
  \nonumber\\
  &=& -\frac{1}{24}\Pi^{[\nu\rho\sigma]}_{[\nu\rho\sigma]}
  \,.
  \label{eq::proj}
\end{eqnarray}
In the practical calculation we consider the object 
$\Pi^{[\nu\rho\sigma]}_{[\nu\rho\sigma]}$ for which we also perform 
the renormalization as described in the following.
Thus, in Eq.~(\ref{eq::proj}) the limit $D\to4$ has been considered
where $D=4-2\epsilon$ is the space-time dimension.

The additional finite counterterm is only needed to one-loop order,
since the singlet diagrams appear the first time at three-loop level.
For each axial-vector vertex a factor~\cite{Trueman:1979en,Larin:1993tq}
\begin{eqnarray}
  Z_5^s &=& 1 - C_F \frac{\alpha_s}{\pi} + {\cal O}(\alpha_s^2)
  \,,
\end{eqnarray}
with $C_F=(N_c^2-1)/(2N_c)$ has to be considered. 
Furthermore, we have to consider the one-loop counterterms for the
strong coupling constant and the top-quark mass defined by
\begin{eqnarray}
  \alpha_s^0 &=& Z_{\alpha_s} \alpha_s
  \,,\qquad
  m_t^0 \,\,=\,\, Z_m m_t
  \,,
\end{eqnarray}
where $m_t\equiv m_t(\mu)$ is renormalized in the
$\overline{\rm MS}$ scheme. The renormalization constants are 
given by
\begin{eqnarray}
  Z_{\alpha_s} &=& 1 + \frac{1}{\epsilon}
  \left(-\frac{11}{12} C_A + \frac{1}{3} T n_f \right)\frac{\alpha_s}{\pi}
  + {\cal O}(\alpha_s^2)  \,,
  \nonumber\\
  Z_m &=& 1 -\frac{3}{4\epsilon} C_F \frac{\alpha_s}{\pi}
  + {\cal O}(\alpha_s^2)  \,,
\end{eqnarray}
with $C_A=N_c$ and $T=1/2$. $n_f=6$ is the number of active flavours.
The transition to the pole mass is achieved via
\begin{eqnarray}
  m_t(\mu) = \left[ 1 +  C_F
    \left(-1-\frac{3}{4}\ln\frac{\mu^2}{M_t^2}\right)\frac{\alpha_s}{\pi}
  + {\cal O}(\alpha_s^2)
  \right] M_t
  \,.
  \label{eq::ms2os}
\end{eqnarray}

We generate the Feynman diagrams with {\tt
  QGRAF}~\cite{Nogueira:1991ex} 
and adopt with
the help of the packages {\tt q2e} and {\tt
  exp}~\cite{Seidensticker:1999bb,Harlander:1997zb}
the topologies and notation to the program performing the reduction
of the four-loop vacuum diagrams~\cite{Schroder:2002re}. As an output we obtain
the corrections to the $\rho$ parameter as a linear combination
of several master integrals. All of them have been computed in
Ref.~\cite{Schroder:2005va}. 

It is interesting to note that some of the master integrals
are multiplied by spurious poles of order $1/\epsilon^2$. 
As a consequence, for these the ${\cal O}(\epsilon)$
and even the ${\cal O}(\epsilon^2)$ contribution is needed.
In the case of the master integral {\tt BB4} 
(which is the four-loop sunset vacuum bubble with one massless
and four massive lines, 
see Eqs.~(4.8) and (6.36) of Ref.~\cite{Schroder:2005va}) it happens that
the coefficient of order $\epsilon$, which originally has only been
evaluated in numerical form~\cite{Schroder:2005va}, enters the pole part
of $\delta\rho$. Thus an analytical expression can be deduced
(in Eq.~(\ref{eq::BB4}) below denoted by ${\tt BB4}^{(1)}$)
which perfectly agrees with the known numerical result.
Furthermore, we have obtained an analytical expression for the coefficient 
of order $\epsilon^2$, by combining the numerically known value with 
the basis of transcendentals known from an independent 
investigation~\cite{Mastrolia:thesis}.
These two coefficients read
\begin{eqnarray}
  {\tt BB4} &=& J^4 \sum_{n\ge0} \epsilon^n\, {\tt BB4}^{(n-4)}
  \,,\nonumber\\
  {\tt BB4}^{(1)} &=&  -\frac{1976975}{7776} + \frac{1792}{9}\zeta(3)
  \nonumber\\
  &=& -14.897726533029588869214274870082319534267\ldots \,, 
  \nonumber\\
  {\tt BB4}^{(2)} &=& 
  - \frac{72443143}{46656} 
  + \frac{47488}{27}\zeta(3)
  - \frac{8704}{3}\zeta(4) 
  \nonumber\\&&\mbox{}
  + \frac{1024}{9}\ln^4 2
  - \frac{2048}{3}\zeta(2)\ln^2 2 
  + \frac{8192}{3}a_4 \nonumber \\ 
  &=& -1678.886929107772963403030310267917509151\ldots \,, 
  \label{eq::BB4}
\end{eqnarray} 
where $J$ is the one-loop tadpole,
$\zeta(n)$ is Riemann's zeta function and
\begin{eqnarray}
  a_4&=&\mbox{Li}_4(1/2)\,\,\approx\,\, 0.51747906167389938633\,.
\end{eqnarray}

Let us mention that we performed the calculation 
using an arbitrary gauge parameter of the
QCD gluon propagator, $\xi$. As expected the final result is independent of 
$\xi$ even before inserting the values for the master integrals.
This constitutes a nice check of our result.


\section{Results and discussion}

Let us in the following present our analytical result and discuss its
numerical implications. For completeness we also repeat the
QCD corrections up to three-loop order.
For the ${\overline{\rm MS}}$ definition of the top-quark mass 
we obtain
\begin{eqnarray}
  \delta\rho^{\overline{\rm MS}} &=& 3 x_t\Bigg\{
  1 + \frac{\alpha_s}{4\pi}\Bigg[
  8 - \frac{16}{3}\zeta(2) + 8\ln\frac{\mu^2}{m_t^2}
  \Bigg]
  + \left(\frac{\alpha_s}{4\pi}\right)^2\Bigg[
  \frac{26459}{81}
  - \frac{25064}{81}\zeta(2) 
  - \frac{3560}{27}\zeta(3)
  \nonumber\\&&\mbox{}
  + \frac{1144}{9}\zeta(4) 
  - \frac{16}{9} B_4 
  - \frac{8}{9}D_3
  + 882 S_2 
  + n_f\left(
    - \frac{50}{3} 
    + \frac{112}{9} \zeta(2) 
    - \frac{64}{9}\zeta(3)
  \right)
  \nonumber\\&&\mbox{}
  - 56\zeta(3) 
  + \left( 
    \frac{668}{3} 
    - \frac{304}{3}\zeta(2)
    + n_f\left(
      - \frac{88}{9} 
      + \frac{32}{9} \zeta(2)
    \right)
  \right) \ln\frac{\mu^2}{m_t^2}
  \nonumber\\&&\mbox{}
  + \left(
    76 - \frac{8}{3} n_f 
  \right) \ln^2\frac{\mu^2}{m_t^2}
  \Bigg]
  + \left(\frac{\alpha_s}{4\pi}\right)^3
  \Bigg[
    \frac{256}{9}
    - 4528\zeta(3)
    + \frac{20816}{3}\zeta(4)
    \nonumber\\&&\mbox{}
    - \frac{2624}{9}\ln^4 2
    + \frac{5248}{3}\zeta(2)\ln^2 2
    - \frac{20992}{3} a_4
    - 1232 \zeta(3) \ln\frac{\mu^2}{m_t^2}
    \Bigg]
  + \ldots
  \Bigg\}
  \label{eq::delrhoMS}
  \,,
\end{eqnarray} 
where the ``$-56\zeta(3)$'' in the third line stems from the 
three-loop singlet diagram. In order $\alpha_s^3$ only the 
singlet contribution is presented. Furthermore, we have
\begin{eqnarray*}
  x_t&=&\frac{G_Fm_t^2}{8\pi^2\sqrt{2}}\,,\\
  S_2&=&\frac{4}{9\sqrt{3}}\mbox{Im}(\mbox{Li}_2(e^{i\pi/3}))
  \,\,\approx\,\,0.26043413763216209896\,,\\
  B_4&=&16 a_4 + \frac{2}{3}\ln^4 2 -4\zeta(2)\ln^2 2 -
  \frac{13}{2}\zeta(4)
  \,\,\approx\,\,-1.7628000870737708641\,,\\
  D_3&=&6\zeta(3) -\frac{15}{4}\zeta(4)  -6
  [\mbox{Im}(\mbox{Li}_2(e^{i\pi/3}))]^2
  \,\,\approx\,\,-3.0270094939876520198\,.
\end{eqnarray*}
In Appendix~\ref{app::res}, we present the three- and four-loop result for
the singlet contribution corresponding to Eq.~(\ref{eq::delrhoMS})
retaining, however, the colour factors $C_F$, $C_A$ and $T$.
With the help of Eq.~(\ref{eq::ms2os}) one obtains the singlet result
in the on-shell scheme. Together with the non-singlet terms one gets
\begin{eqnarray}
  \delta\rho^{\rm OS} &=& 3 X_t\Bigg\{
  1 + \frac{\alpha_s}{4\pi}\Bigg[
  -\frac{8}{3} - \frac{16}{3}\zeta(2)
  \Bigg]
  + \left(\frac{\alpha_s}{4\pi}\right)^2\Bigg[
  \frac{314}{81}
  - \frac{26504}{81}\zeta(2) 
  - \frac{3416}{27}\zeta(3)
  \nonumber\\&&\mbox{}
  - \frac{64}{3} \zeta(2) \ln2 
  + \frac{1144}{9}\zeta(4) 
  - \frac{16}{9} B_4 
  - \frac{8}{9}D_3
  + 882 S_2 
  + n_f\left(
    - \frac{8}{9} 
    - \frac{208}{9} \zeta(2) 
    \right.\nonumber\\&&\left.\mbox{}
    - \frac{64}{9}\zeta(3)
  \right)
  - 56\zeta(3) 
  + \left( 
    - \frac{88}{3} 
    - \frac{176}{3}\zeta(2)
    + n_f\left(
        \frac{16}{9} 
      + \frac{32}{9} \zeta(2)
    \right)
  \right) \ln\frac{\mu^2}{M_t^2}
  \Bigg]
  \nonumber\\&&\mbox{}
  + \left(\frac{\alpha_s}{4\pi}\right)^3
  \Bigg[
    \frac{256}{9}
    - \frac{11792}{3}\zeta(3)
    + \frac{20816}{3}\zeta(4)
    \nonumber\\&&\mbox{}
    - \frac{2624}{9}\ln^4 2
    + \frac{5248}{3}\zeta(2)\ln^2 2
    - \frac{20992}{3} a_4
    - 784 \zeta(3) \ln\frac{\mu^2}{M_t^2}
    \Bigg]
  + \ldots
  \Bigg\}
  \label{eq::delrhoOS}
  \,,
\end{eqnarray} 
with $X_t=G_FM_t^2/(8\pi^2\sqrt{2})$.

Inserting the numerical values for the constants in 
Eqs.~(\ref{eq::delrhoMS}) and~(\ref{eq::delrhoOS}) 
and adopting $\mu=m_t$ and $\mu=M_t$, respectively, the 
numerical corrections read
\begin{eqnarray}
  \delta\rho^{\overline{\rm MS}} &=& 3 x_t\Bigg[
  1 -0.19325 \frac{\alpha_s}{\pi} +  
  ( - 4.2072 + 0.23764 ) \left(\frac{\alpha_s}{\pi}\right)^2    
  - 3.2866 \left(\frac{\alpha_s}{\pi}\right)^3
  \Bigg]
  \,,
  \nonumber\\
  \delta\rho^{\rm OS} &=&
  \label{eq::delrhoOSMSnum}
  3 X_t\Bigg[
  1 - 2.8599 \frac{\alpha_s}{\pi} +
  (- 4.2072 -10.387) \left(\frac{\alpha_s}{\pi}\right)^2
  + 7.9326 \left(\frac{\alpha_s}{\pi}\right)^3
  \Bigg]
  \,,
\end{eqnarray} 
where the three-loop contribution is split into the 
singlet (first number in round brackets) and the non-singlet piece.
If we furthermore adopt $\alpha_s(m_t)=0.108$ and
$\alpha_s(M_t)=0.107$, the expression for $\delta\rho$ looks like
\begin{eqnarray}
  \delta\rho^{\overline{\rm MS}} &=& 3 x_t\Bigg(
  1 - 0.00664 - 0.00469 - 0.00013
\Bigg)
  \,,
  \nonumber\\
  \delta\rho^{\rm OS} &=& 3 X_t\Bigg(
  1 - 0.09741 - 0.01693 + 0.00031
  \Bigg)
  \label{eq::delrhoOSMSnum2}
  \,,
\end{eqnarray} 
where the n$^{\rm th}$ term inside the round brackets corresponds
to the contribution of order $G_F M_t^2 \alpha_s^{(n-1)}$.
One observes that the new four-loop singlet contribution is
numerically small and amounts to
about 3\% of the three-loop result in the $\overline{\rm MS}$ scheme
and to less than 2\% for on-shell top-quark masses.
Note that the correction is positive in the on-shell and negative in
the $\overline{\rm MS}$ scheme. 
In the on-shell scheme the shift in $M_W$ and $\sineff$ according to
Eqs.~(\ref{eq::delMW}) and~(\ref{eq::delsin})
amounts to 0.175~MeV and $10^{-6}$, respectively,
which is significantly below the recent estimates
of higher order contributions and variations of input 
parameters~\cite{Awramik:2003rn,Awramik:2004ge}.

It is interesting to mention that at three-loop order the singlet contribution 
completely dominates for $\overline{\rm MS}$ top-quark masses and
amounts to about 30\% in the on-shell scheme. Thus, in case the
same pattern also holds at four-loop order, the complete QCD corrections
would be well under control.
However, the numerical values in Eq.~(\ref{eq::delrhoOSMSnum}) suggest
that for some reason the four-loop singlet contribution seems to be
accidentally small.

Let us also comment on the dependence of the singlet
contribution on the renormalization scale $\mu$ which can be done
separately from the non-singlet part. The latter is discussed in
Ref.~\cite{Chetyrkin:1995ix} (cf. Fig.~1 of
Ref.~\cite{Chetyrkin:1995ix}). As far as the singlet contribution is 
concerned one obtains for the quantity $(\delta\rho^{\rm
OS}/(3X_t)-1)_{\rm singlet}$ the values $\{-0.00457,-0.00437,-0.00455\}$
corresponding to $\mu=\{M_t,M_t/2,2M_t\}$.
The $\mu$-dependence, being formally of higher order, 
is less then 5\% of the sum of the three- and four-loop singlet 
part which can be used as an estimate of the 
${\cal O}(\alpha_s^4)$ term.

In the remaining part of this section we briefly compare 
the numerical effect of the new terms with known corrections to $\delta\rho$.
In the on-shell scheme the three-loop QCD corrections 
of order $\alpha_s^2 X_t$ lead to a shift
of about $-10$~MeV in the $W$-boson mass and to $+5\times 10^{-5}$ in
the effective weak mixing angle. For Higgs-boson masses between
200~GeV and 300~GeV the three-loop corrections of order 
$\alpha_s X_t^2$~\cite{Faisst:2003px} have the opposite sign 
and with roughly half the magnitude they are still relevant for
the precision to be reached at the Giga-$Z$ option of a future $e^+e^-$
linear collider. 
However, the pure electroweak corrections of order $X_t^3$ are very
small and give rise to corrections well below 1~MeV for the shift
in the $W$-boson mass.
The same is true for the four-loop QCD singlet contributions
considered in this letter.

In conclusion, we computed the four-loop singlet contribution to the
$\rho$ parameter which constitutes one of the first applications of the
four-loop massive vacuum integrals to a physical quantity.
The numerical size of the corrections turn out to be surprisingly small
and lead to a shift in the $W$-boson mass below 1~MeV
and to the effective weak mixing angle below $10^{-5}$
--- beyond the accuracy forseen in a future Linear Collider.
This illustrates the good convergence properties of the perturbation
theory and confirms the stable predictions based on the three-loop
corrections.
However, for a definite conclusion also the non-singlet contribution
has to be evaluated.


\vspace*{1em}

\noindent
{\bf Acknowledgements}\\
This work was supported by SFB/TR 9. We thank K.G. Chetyrkin and
J.H. K\"uhn for carefully reading the manuscript.


\begin{appendix}

\section{\label{app::res}Singlet contribution to the $\rho$ parameter}

In this Appendix we present the three- and four-loop singlet result
expressed in terms of $C_A=N_c$, $C_F=(N_c^2-1)/(2N_c)$ and $T=1/2$.
Furthermore, we keep the label $n_l$ which counts the number of massless
quarks. The three-loop term can also be found in Ref.~\cite{Anselm:1993uq}.
\begin{eqnarray}
  \delta\rho^{\overline{\rm MS}}_{\rm sing} &=& 3 x_t
  \left(\frac{\alpha_s}{4\pi}\right)^2 C_F T 
  \Bigg\{
  -84\zeta(3) 
  + \frac{\alpha_s}{4\pi} \Bigg[
    C_F \Bigg(
  - 336\zeta(3) 
  + 2400\zeta(4)
  - 128\ln^4 2 
  \nonumber\\&&\mbox{}
  + 768\zeta(2)\ln^2 2 
  -3072 a_4 
  - 504\zeta(3)\ln\frac{\mu^2}{m_t^2}
  \Bigg)
  + C_A \left(
  - \frac{7064}{3}\zeta(3)
  \right.\nonumber\\&&\left.\mbox{}
  + 3056\zeta(4)
  - \frac{320}{3}\ln^4 2 
  + 640\zeta(2)\ln^2 2 
  - 2560 a_4 
  - 616\zeta(3)\ln\frac{\mu^2}{m_t^2}
  \right)
  \nonumber\\&&\mbox{}
  + n_l T \left( 
    \frac{1120}{3}\zeta(3)
  - 784\zeta(4)
  + \frac{64}{3}\ln^4 2 
  - 128\zeta(2)\ln^2 2 
  + 512 a_4 
  \right.\nonumber\\&&\left.\mbox{}
  + 224\zeta(3)\ln\frac{\mu^2}{m_t^2}
  \right)
  + T \left( 
  \frac{256}{3} 
  - \frac{1280}{3}\zeta(3) 
  + 224\zeta(3)\ln\frac{\mu^2}{m_t^2}
  \right)
  \Bigg]
  + {\cal O}(\alpha_s^2)
  \Bigg\}
  \,.
  \nonumber\\
\end{eqnarray}

\end{appendix}




\begin{thebibliography}{99}


\bibitem{Veltman:1977kh}
  M.~J.~G.~Veltman,
  Nucl.\ Phys.\ B {\bf 123} (1977) 89.

\bibitem{Sirlin:1980nh}
  A.~Sirlin,
  Phys.\ Rev.\ D {\bf 22} (1980) 971.

\bibitem{LEPEWWG}
See, e.g., {\tt http://lepewwg.web.cern.ch/LEPEWWG/Welcome.html}.

\bibitem{Aguilar-Saavedra:2001rg}
  J.~A.~Aguilar-Saavedra {\it et al.}  [ECFA/DESY LC Physics Working
  Group],
  hep-ph/0106315.

\bibitem{Djouadi:1987gn}
  A.~Djouadi and C.~Verzegnassi,
  Phys.\ Lett.\ B {\bf 195} (1987) 265.

\bibitem{Djouadi:1987di}
  A.~Djouadi,
  Nuovo Cim.\ A {\bf 100} (1988) 357.

\bibitem{Kniehl:1988ie}
  B.~A.~Kniehl, J.~H.~K\"uhn and R.~G.~Stuart,
  Phys.\ Lett.\ B {\bf 214} (1988) 621.
 
\bibitem{Avdeev:1994db}
  L.~Avdeev, J.~Fleischer, S.~Mikhailov and O.~Tarasov,
  Phys.\ Lett.\ B {\bf 336} (1994) 560
  [Erratum-ibid.\ B {\bf 349} (1995) 597]
  [hep-ph/9406363].

\bibitem{Chetyrkin:1995ix}
  K.~G.~Chetyrkin, J.~H.~K\"uhn and M.~Steinhauser,
  Phys.\ Lett.\ B {\bf 351} (1995) 331
  [hep-ph/9502291].

\bibitem{vanderBij:1986hy}
  J.~J.~van der Bij and F.~Hoogeveen,
  Nucl.\ Phys.\ B {\bf 283} (1987) 477.

\bibitem{Barbieri:1992nz}
  R.~Barbieri, M.~Beccaria, P.~Ciafaloni, G.~Curci and A.~Vicere,
  Phys.\ Lett.\ B {\bf 288} (1992) 95
  [Erratum-ibid.\ B {\bf 312} (1993) 511]
  [hep-ph/9205238];\\
  J.~Fleischer, O.~V.~Tarasov and F.~Jegerlehner,
  Phys.\ Lett.\ B {\bf 319} (1993) 249.



\bibitem{Faisst:2003px}
  M.~Faisst, J.~H.~K\"uhn, T.~Seidensticker and O.~Veretin,
  Nucl.\ Phys.\ B {\bf 665} (2003) 649
  [hep-ph/0302275].


\bibitem{Boughezal:2004ef}
  R.~Boughezal, J.~B.~Tausk and J.~J.~van der Bij,
  Nucl.\ Phys.\ B {\bf 713} (2005) 278
  [hep-ph/0410216].

\bibitem{Boughezal:2005eb}
  R.~Boughezal, J.~B.~Tausk and J.~J.~van der Bij,
  hep-ph/0504092.

\bibitem{Awramik:2003rn}
  M.~Awramik, M.~Czakon, A.~Freitas and G.~Weiglein,
  Phys.\ Rev.\ D {\bf 69} (2004) 053006
  [hep-ph/0311148].

\bibitem{Awramik:2004ge}
  M.~Awramik, M.~Czakon, A.~Freitas and G.~Weiglein,
  Phys.\ Rev.\ Lett.\  {\bf 93} (2004) 201805
  [hep-ph/0407317].

\bibitem{Schroder:2005va}
  Y.~Schroder and A.~Vuorinen,
  JHEP {\bf 0506} (2005) 051
  [arXiv:hep-ph/0503209].

\bibitem{'tHooft:1972fi}
  G.~'t Hooft and M.~J.~G.~Veltman,
  Nucl.\ Phys.\ B {\bf 44} (1972) 189.

\bibitem{Larin:1993tq}
  S.~A.~Larin,
  Phys.\ Lett.\ B {\bf 303} (1993) 113
  [hep-ph/9302240].

\bibitem{Trueman:1979en}
  T.~L.~Trueman,
  Phys.\ Lett.\ B {\bf 88} (1979) 331.

\bibitem{Nogueira:1991ex}
  P.~Nogueira,
  J.\ Comput.\ Phys.\  {\bf 105} (1993) 279.

\bibitem{Seidensticker:1999bb}
  T.~Seidensticker,
  hep-ph/9905298.

\bibitem{Harlander:1997zb}
  R.~Harlander, T.~Seidensticker and M.~Steinhauser,
  Phys.\ Lett.\ B {\bf 426} (1998) 125
  [hep-ph/9712228].

\bibitem{Schroder:2002re}
  Y.~Schr\"oder,
  Nucl.\ Phys.\ Proc.\ Suppl.\  {\bf 116} (2003) 402
  [hep-ph/0211288].

\bibitem{Mastrolia:thesis}
  P.~Mastrolia, 
  PhD thesis (May 2004), Bologna Univ., Italy (unpublished).

\bibitem{Anselm:1993uq}
  A.~Anselm, N.~Dombey and E.~Leader,
  Phys.\ Lett.\ B {\bf 312} (1993) 232.

\end{thebibliography}
\end{document}